\documentclass{article}
\usepackage{hyperref} 
\hypersetup{
    colorlinks=true,
    linkcolor=blue,
    urlcolor=blue,
}

\usepackage{graphicx}%
\usepackage{multirow}%
\usepackage{amsmath,amssymb,amsfonts}%
\usepackage{amsthm}%
\usepackage{mathrsfs}%
\usepackage[title]{appendix}%
\usepackage{xcolor}%
\usepackage{textcomp}%
\usepackage{manyfoot}%
\usepackage{booktabs}%
\usepackage{algorithm}%
\usepackage{algorithmic}
\usepackage{listings}%
\usepackage{soul}
\usepackage{authblk} 

\title{Chain-structured neural architecture search for financial time series forecasting}

\author[1]{Denis Levchenko\thanks{denis.levchenko@heig-vd.ch}}
\author[1]{Efstratios Rappos}
\author[1]{Shabnam Ataee}
\author[2]{Biagio Nigro}
\author[1]{Stephan Robert-Nicoud}

\affil[1]{School of Engineering and Management Vaud (HEIG-VD), University of Applied Sciences and Arts Western Switzerland (HES-SO), Yverdon-les-Bains, Switzerland}
\affil[2]{Predictive Layer SA, Rolle, Switzerland}

\date{} 

\begin{document}

\maketitle

\begin{abstract}
    Neural architecture search (NAS) emerged as a way to automatically optimize neural
    networks for a specific task and dataset. Despite an abundance of research on NAS for
    images and natural language applications, similar studies for time series data are lacking.
    Among NAS search spaces, chain-structured are the simplest and most applicable to small datasets
    like time series. We compare three popular NAS strategies on
    chain-structured search spaces: Bayesian optimization (specifically
    Tree-structured Parzen Estimator), the hyperband method,
    and reinforcement learning in the context of financial time series forecasting. These strategies
    were employed to optimize simple well-understood neural architectures like the MLP, 1D CNN, and
    RNN, with more complex temporal fusion transformers (TFT) and their own optimizers included
    for comparison.
    We find Bayesian optimization and the hyperband method performing best among the strategies,
    and RNN and 1D CNN best among the architectures, but all methods were very close to
    each other with a high variance due to the difficulty of working with financial datasets.
    We discuss our approach to overcome the variance and provide implementation recommendations
    for future users and researchers.
\end{abstract}

\vspace{1em} 

\vspace{1em} 
\noindent This is the accepted version of the paper published in
\emph{International Journal of Data Science and Analytics}. \\
DOI: \href{https://doi.org/10.1007/s41060-024-00690-y}{10.1007/s41060-024-00690-y}

\section{Introduction}\label{sec:Introduction}

Deep neural networks have been very successful in a wide variety of tasks over
the last two decades. In large part their success is attributed to their
ability to perform very well without major manual feature engineering
required when
compared to more classical techniques.~\cite{GoodfellowDeep}.
However, the exact architecture of the neural
network still has to be prescribed manually by the user. This led to the
development of so-called auto-ML techniques that aim to automate this process.
In the context of deep neural networks, auto-ML has a very large overlap with
neural architecture search (NAS), itself having a large overlap with
hyperparameter optimization. 

A lot of research
has been done in NAS in recent years, see~\cite{White_NAS_survey} for an
overview and insights from over 1000 papers. However, most work focused on
computer vision or natural language applications, with less investigation
into architectures for analyzing time series data.
In this work, we attempt to bridge
this gap by evaluating and comparing the performance of three popular
simple NAS strategies on 3 distinct yet similar (financial) time series
datasets.

Modern competitive neural nets perform very well on natural language and
images datasets, and any improvement is marginal, while
the datasets and networks are huge, requiring vast
resources to train each individual network and thus directing NAS research
toward more complicated methods that reduce the number of overall
trials~\cite{White_NAS_survey, ElskenNAS}.
Meanwhile, time series, and especially financial time series,x pose a
fundamentally different challenge.
With time series, the datasets are often small due to the low number of
historical samples, which forces the networks to have fewer parameters too
in order to avoid overfitting. This results in neural networks that train
in seconds on modern hardware. The performance, especially in the case of
financial time series, on the other hand, is always
comparatively very poor. The challenge comes with financial markets being
notoriously hard to
predict. Intuitively, this is clear: if it was easy, every researcher
would have been a millionaire.
The problem is challenging due to the inherently
noisy and non-stationary nature of financial time series. Market prices are
influenced by a myriad of unpredictable factors, such as economic events,
investor behavior, and geopolitical developments, leading to a low
signal-to-noise ratio. Consequently, even state-of-the-art algorithms struggle
to achieve high performance, typically resulting in F1, (balanced) accuracy,
and AUC
scores well below 0.6 on average when applied to test
data~\cite{LSTMForFinance, CNNpred, ClassificationFinanceNN}.

A seminal survey article on NAS~\cite{ElskenNAS} breaks down the search into
3 components: search spaces, search strategies, and performance estimation
strategies. Once a search space is set up, an algorithm following the search
strategy explores the search space looking for the best neural network
configuration, evaluated and
eventually selected by the performance estimation strategy. Following this
terminology, we discuss the
search spaces we set up for NAS in Section~\ref{sec:Architectures}, the
search strategies in Section~\ref{sec:Strategies} and the performance
estimation together with methodology in
Section~\ref{sec:Estimation}. In
Section~\ref{sec:Problem} we discuss the exact practical problem
from project partner Predictive Layer SA that our neural
networks were trying to solve. Our results are summarized and discussed in
Section~\ref{sec:Results} and Section~\ref{sec:Discussion}, respectively.
Finally, we outline possible future directions for this research
in Section~\ref{sec:Future}.

\section{Data and problem formulation}
\label{sec:Problem}
This project was a collaboration with Predictive Layer SA, who provided
real-world customer data and prediction requirement. 
The data under consideration are financial multivariate daily time series
in tabular form, each row corresponding to one day and each
column a separate feature. The task consists of predicting whether the target
feature will increase or decrease 5 days (or 10 days, depending on the dataset)
in the future.
Thus the problem is essentially
binary classification. The models output a number between $0$ and $1$ as output,
interpreted as probability that the target feature will grow on a given day.

Our three source datasets that we report on in
Section~\ref{sec:Results} are for Japanese, German, and US bonds. Each of
the tabular source datasets has a size of about $4000\times1000$: $1000$
columns (input
features), representing the many financial markers believed to be
predictive of the target, with each row corresponding to a business day
over a 15-year period. While this sample size is relatively small for
deep learning applications~\cite{GoodfellowDeep} -- especially given the
high number of features -- such sizes are typical in time series datasets.
Thus we hope that our findings will be relevant to time series practitioners.

\section{Architecture types and their search spaces}
\label{sec:Architectures}
In NAS, the search spaces
and search strategies go hand in hand and are designed together essentially
at the same time. The
simplest among the search spaces are chain-structured. Search spaces
with a chain structure feature a straightforward
architectural topology: a sequential chain of operational
layers. These configurations frequently utilize cutting-edge manually designed
architectures as their foundational framework, trying to find the best
configuration for the specific dataset at hand by essentially hyperparameter
optimization. Despite their simplicity, chain-structured search spaces
often yield very good results~\cite{White_NAS_survey}.
The biggest downsides of this approach when
compared to the more complicated cell-based and one-shot approaches are its
limited flexibility and scalability when applied to very large datasets.
While being limited to a pre-defined overall structure prevents discovering
truly novel architectures, in practice well-designed and well-understood
architectures usually perform best~\cite{White_NAS_survey}. Scalability becomes
important when dealing with large image and natural language datasets that most
research on NAS focuses on; however, in the case of time series and especially
our small datasets, our neural networks train in seconds. Therefore scalability
was not an issue for us, and thus we focused on chain-structured search spaces.

For the foundational
frameworks that the search spaces were built upon, we concentrated on simple
well-known architecture types, like feedforward (FFNN, also known as
multi-layer perceptron or MLP), convolutional (CNN), and recurrent (RNN) networks.
This was done both for their proven performance in time-series context and
due to them having fewer parameters than most cutting-edge architectures, which
is more appropriate when dealing with very little data as in our case.
However, we also examined the state of the art Temporal Fusion
Transformer (TFT) architecture for comparison (see below).
Each of these architectures
was adapted to the problem at hand and for each of them a search space of
hyperparameters was set up to determine the
exact configuration. The search strategies discussed in
Section~\ref{sec:Strategies} then optimized each of these separately, and
the best-performing ones were
compared. This approach is similar to what \cite{Zoph} did with a single
type of CNN for image classification.

\subsection{Feedforward networks}
Feedforward networks (aka multi-layer perceptron or MLP) have the advantage
of being the simplest type of deep
neural networks. The main disadvantage for our problem comes from the fact
that they cannot take the time dimension into account explicitly. Feedforward
networks take a row of numbers as input, one row of our datasets
corresponding to one day information for various time series. Implicitly, the
time information is taken into account in the form of delayed (or averaged)
features -- a
column in the dataset that is essentially the same as another column with
an added delay of a couple days (or averaged over a couple days, respectively),
a standard practice in the field~\cite{CNNpred}.

As far as NAS is concerned, our hyperparameters for optimization are the number
of hidden layers and the number of units per hidden layer. Dropout layers
were used for regularization and the dropout rate was another
hyperparameter optimized. Although not part of the architecture itself,
the learning rate is a crucial hyperparameter for any learning algorithm
that was also optimized.

\subsection{Convolutional neural networks}
Although designed for and used mainly with images,
convolutional neural networks (CNNs)
are suitable for any kind of grid-like data, especially when there is a natural
fixed notion of distance between neighboring data points, such as the case for
time series~\cite{GoodfellowDeep}.

CNNs allow taking the time-dimension into account explicitly and naturally.
Instead of each input being a single row of data corresponding to a single
business day (as was the case for FFNN), a sliding window or a `chunk' of
our dataset -- a
subtable comprised of several consecutive rows
can be passed as a single input to a CNN. Evaluating the history
from the last, say, $15$ business days explicitly,
the model is making a prediction for the following day as before.

Both one- and two-dimensional CNNs can be used for our problem.
One-dimensional (1D) CNNs only perform the convolution operation
over the time dimension,
treating separate features as parallel input channels,
just like the red, green, and blue channels are considered
when working with images. Thus each chunk of data is passed
as a 1D `image' of size \texttt{chunk\_length} (a hyperparameter to be optimized)
and $n_{features}$ (fixed number coming from the dataset) channels.
Two-dimensional (2D) CNNs
take each chunk as a single-channel 2D `image' of size
\texttt{chunk\_length}~$\times$~$n_{features}$, performing the convolution
operation over both the time dimension and across the various features. The
latter seems counterintuitive as there is no fixed distance between the
features, but such approach was found to be successful in the context of
financial time series before\cite{CNNpred}. After
experimenting with 2D CNNs on our data, we found them a lot more
computationally expensive than others, while often performing worse. For
this reason 2D CNNs were later dropped from our consideration.

The length in time of the chunks passed as individual examples,
\texttt{chunk\_length},
was a crucial hyperparameter to optimize for
all of the CNNs. The number of convolutional layers (including pooling and
activation functions) was fixed at $3$ for all CNN architectures to reduce the
size of the search spaces, but the
kernel sizes were optimized. The number of convolution filters per layer was
optimized too, but it works differently for different implementations,
as discussed below. Dropout layers were used once again for
regularization and the dropout rate was optimized too, together with
the learning rate.

Among 1D CNNs, there are two distinct ways to perform
the convolution operation. In \emph{depthwise} convolution, each channel
(input feature) is passed separately through its own convolutional and
pooling layers.
Each of these then gives a fixed number (the number of separate convolution
filters) of output channels per input channel,
interpreted as a summary of what happened to this feature in the
time span considered. The convolution and pooling operations will shrink the
length (in time) of each output channel, possibly to $1$ (otherwise
flattening is applied), and all of these are then passed to a
fully connected layer that makes a prediction based on these summaries.
Although easy to interpret, this method had the problem of blowing up the
dimensionality of the problem, as each of the features gets its own set of
convolution parameters. This made computation infeasible and we abandoned
this approach.
In basic (we call them `ordinary') 1D convolution, all input channels are
mixed already in the first (and subsequent) convolution layers. The output of
convolution is
flattened and passed to a single fully connected layer for making the final
prediction.

\subsection{Recurrent neural networks}
Recurrent neural networks (RNNs) were designed for and are very successful at
treating sequential data~\cite{GoodfellowDeep}
and are thus the natural choice for our problem. The same chunks of data as
for CNNs were passed as individual inputs to the RNNs, here naturally
interpreted as sequences of length \texttt{chunk\_length} and dimension
\texttt{n\_features}. Both simple RNNs and
more sophisticated long short-term memory (LSTM) RNNs were implemented and
optimized separately, but they have the same search spaces of hyperparameters.
In both cases, \emph{stacked} RNNs were used, where the output of one RNN is
taken as input to the next one. The number of such stacked layers was a
hyperparameter to optimize. Each RNN layer also has its number of hidden
units, which was also optimized. The chunk length was once again
optimized, together with dropout and learning rates as before.

LSTMs provide many advantages over basic RNNs~\cite{GoodfellowDeep} without
too much additional computational cost, especially in the case of our small datasets.
For this reason after initial trials, we proceeded with
just LSTMs for the final tests.

\subsection{Temporal fusion transformer}
Forecasting across multiple time horizons often involves an intricate
amalgamation of inputs, encompassing static (i.e., time-invariant)
covariates, known future inputs, and additional exogenous time series
observed solely in the past. The challenge lies in handling this complexity
without prior knowledge of how these inputs interact with the target variable.
While various deep learning approaches have been proposed, they often manifest
as `black-box' models, lacking transparency regarding their utilization of the
diverse input types encountered in practical scenarios.

In a recent work~\cite{TFT}, the Temporal Fusion Transformer (TFT) was introduced
as an innovative attention-based architecture addressing this issue. TFT not
only achieves high-performance multi-horizon forecasting but also provides
interpretable insights into temporal dynamics. In order to capture temporal
relationships
at different scales, TFT incorporates recurrent layers for local processing and
interpretable self-attention layers for modeling long-term dependencies.
Specialized components within TFT are employed to select relevant features, and
a series of gating layers effectively suppress unnecessary components,
resulting in high performance across a broad spectrum of scenarios.

~\cite{TFT} demonstrated significant performance enhancements over existing
benchmarks across various real-world datasets. The study also highlighted three
practical use cases illustrating the interpretability of TFT, showcasing its
efficacy in shedding light on the decision-making process. The PyTorch
Forecasting package~\cite{PyTorchForecasting}
has on open-source implementation
of the model. It comes with its own optimizer, selecting the optimal number
of attention heads, the network's hidden size, and learning and dropout rates.

Although very promising and successful for other datasets, we found performance
of TFT rather poor on our data: the models reduce to a trivial binary predictor,
always predicting $1$ or $0$ no matter the input test data. We believe this was
largely due to the lack of data. As previously mentioned, novel complex
architectures like TFT have tens if not hundreds of thousands of parameters
which need a lot
of data to train well. It would be interesting to revisit TFT and other models
derived from it that have recently appeared in the literature on other, bigger
datasets.

\section{Search strategies}
\label{sec:Strategies}
Classical approaches to hyperparameter tuning in machine learning, such as
grid search and random search, have been widely used due to their simplicity
\cite{BergstraRandom}. However, these methods can be computationally
expensive and inefficient,
especially when dealing with high-dimensional search spaces or expensive
objective functions~\cite{BergstraAlgorithms}. Many approaches have been
introduced to improve performance, including but not limited to: particle
swarm optimization~\cite{PSO}, using a simple neural
predictor~\cite{NeuralPredictorNAS}, NAS without training~\cite{NASNoTraining},
Bayesian optimization, reinforcement
learning, and the hyperband method.
An extensive survey on over 1000 NAS papers~\cite{White_NAS_survey} favorably
featured the three latter approaches among the strategies for chain-structured
search spaces, and these are the methods we explored
in our work. The survey also prioritized cell-based, hierarchical search spaces
and one-shot techniques, but we believe chain-structured spaces make more sense
for small datasets like time series.

\subsection{Bayesian optimization}

Bayesian optimization has emerged as a popular alternative to grid and random
search for hyperparameter tuning, as it efficiently explores the search space
and intelligently guides the optimization process~\cite{SnoekPractical}. One
widely used technique in Bayesian optimization is based on Gaussian
processes (GPs), which fit the objective function using Gaussian processes as
probabilistic models and leverages acquisition functions to balance exploration
and exploitation~\cite{RasmussenGaussian}. GPs provide a flexible,
nonparametric tool for modeling complex functions, allowing for uncertainty
quantification and adaptation to new data~\cite{ShariariTaking}.

However, GP-based methods have some limitations, such as the inability to
handle categorical features or dependent parameters without ad hoc
modifications~\cite{BrochuTutorial}. This has led to the development of
alternative methods, such as the Tree-structured Parzen Estimator (TPE),
which has gained attention for its scalability and
effectiveness~\cite{BergstraAlgorithms}. TPE models the search space using
hierarchical Parzen estimators, which adaptively partition the space to model
complex, high-dimensional functions~\cite{BergstraAlgorithms}. A simplified
high-level version of the TPE algorithm for neural networks optimization is
presented in
Algorithm~\ref{alg:Bayesian}. Line 6 in Algorithm~\ref{alg:Bayesian}
makes intuitive
sense as `choose a point where $x_{next}$ is most likely ``good'' and least
likely ``bad'''. Theoretically, it is justified since optimizing the criterion
of Expected
Improvement (EI) -- the goal of TPE -- is equivalent to
maximizing $l(x)/g(x)$ at each step of the algorithm~\cite{BergstraAlgorithms}.
See~\cite{BergstraAlgorithms} for a more detailed
introduction to TPE and~\cite{WatanabeTPE} for an in-depth tutorial.

\begin{algorithm}
    \caption{Simplified TPE NAS}
    \begin{algorithmic}[1]
    \label{alg:Bayesian}
        \REQUIRE A search space $S$ of neural network
        hyperparameters, number of trials $n_{trials}$, number of initial
        random samples $n_{init}$
        \ENSURE Results $R$ for each hyperparameter configuration tested
        \STATE Randomly select $n_{init}$ configurations from $S$.
        \STATE Train and test neural nets with the selected configurations,
            recording results in $R$
        \FOR{$n=1$ to $n_{trials}$}
            \STATE Split $R$ into $R_{good}$ and $R_{bad}$ based on their
            performance according to the test metric.
            \STATE Fit two density models: $l(x)$ for $R_{good}$ and $g(x)$
            for $R_{bad}$
            \STATE Choose the next configuration $x_{next}$ by maximizing
            $l(x_{next})/g(x_{next})$
            \STATE Train and test a neural net with the new configuration,
            updating $R$
        \ENDFOR
        \RETURN $R$
    \end{algorithmic}
    \end{algorithm}

TPE is particularly suitable for addressing the shortcomings of GP-based
methods, as it can naturally handle categorical features and dependent
parameters without requiring extensive adaptations. Performance comparisons
between TPE, GP-based methods, and classical search techniques have shown that
TPE often yields improved results across various tasks and
datasets~\cite{EggenspergerTowards}.

A disadvantage of Bayesian optimization in general and TPE especially is the
high theoretical complexity of the method when compared with simpler
techniques, often for not much
gain~\cite{White_NAS_survey, NeuralPredictorNAS}.
Thankfully, a good well-supported practical implementation of TPE (that we
used for our comparison) is available
in the \texttt{Optuna} package, a
versatile optimization library that has demonstrated its utility in various
machine learning tasks~\cite{AkibaOptuna}.

\subsection{Reinforcement learning approach}
An alternative iterative approach consists of treating the neural architecture
search problem as a reinforcement learning problem. Action of the agent at time
step $t$ is selecting a specific neural network configuration. The reward is
the performance of the selected model which leads to the next action. In this
way choices of parameters which produce ``good'' neural
networks are rewarded, whereas those which produce badly-performing neural
networks are penalized. We let the system evolve for a number of
iterations and at the end obtain the test results for all configurations
explored. 

The specific method we used introduced in~\cite{Zoph} uses an RNN as a
controller, trained with REINFORCE. The output sequence of the RNN controller
is a string encoding the hyperparameters of a neural network to be trained and
tested. The authors claimed that compared with
Bayesian optimization, the reinforcement learning method is more
general and more flexible~\cite{Zoph}. The algorithm is presented
in~Algorithm~\ref{alg:Reinforcement}.
An open-source implementation of the
controller RNN to use with image data is available at~\cite{ZophImplementation},
which we adapted to work with our time series data.

\begin{algorithm}
\caption{Reinforcement learning NAS}
\begin{algorithmic}[1]
\label{alg:Reinforcement}
    \REQUIRE A search space $S$ of neural network
    hyperparameters, number of trials $n_{trials}$
    \ENSURE Results $R$ for each hyperparameter configuration tested
    \STATE Initialize the controller RNN $C$ with random weights $\theta _C$
    \FOR{$i=1$ to $n_{trials}$}
        \STATE Evaluate $C$ to get an architecture configuration $s$ from $S$
        \STATE Train and test a neural net with configuration $s$,
        update results $R$
        \STATE Use the test metric as the reward signal, compute the policy
        gradient and update the controller weights $\theta _C$ of $C$
    \ENDFOR
    \RETURN $R$
\end{algorithmic}
\end{algorithm}

\subsection[Hyperband]{Hyperband}
The Hyperband introduced in 2018 by Li et al.~\cite{hyperband},
is a hyperparameter optimization method fundamentally different from both
Bayesian optimization and the reinforcement learning approach.
While the latter both
use a clever technique to suggest the next neural network configuration to
fully train and test, the Hyperband method selects many configurations
randomly, instead cleverly allocating resources only to the ones looking
promising early on in their training.

What makes Hyperband notable are its simplicity, efficiency, scalability,
and the ability to use multiple GPUs simultaneously for faster optimization.
It is compatible with popular machine learning libraries like
\texttt{Scikit-learn} and
\texttt{Keras}, embedded in the \texttt{KerasTuner}
framework~\cite{keras_tuner}.

Hyperband is designed for the efficient exploration and optimization of
hyperparameter configurations. 
Algorithm~\ref{alg:Hyperband}
employs \textit{successive halving}, a subprocess that involves eliminating
less promising models and further training the survivors.

In successive halving, you provide the hyperparameter search space $S$,
the total number of epochs $epochs_{\text{max}}$ needed for full training
of a neural net, and a factor $\eta>1$ (usually $\eta=2$ or $3$ or $4$)
determining the speed of
reducing models and increasing training epochs for the survivors.
This process iterates until the best performing models are fully trained,
resulting in the identification of the best model and its validation loss.

To achieve a balance between exploration and optimization, Hyperband executes
successive halving multiple times, labeling each iteration as a
\textit{bracket}. The number of brackets is determined by a parameter
$ s_{\text{max}} = \left\lfloor \log_\eta (epochs_{\text{max}}) \right\rfloor $.
In each bracket,
the number of models decreases by a factor, and new models are generated.
Hyperband ensures a robust exploration by randomly generating models for each
bracket, preventing too few models in a bracket. The initial training epochs for
each bracket strictly increase by a factor.

\begin{algorithm}
    \caption{Hyperband NAS}
    \begin{algorithmic}[1]
    \label{alg:Hyperband}
    \REQUIRE Maximum number of epochs $epochs_{\text{max}}$,
    hyperparameter search space $S$, reduction factor $\eta > 1$
    \ENSURE Set $R$ of best hyperparameters configurations maximizing the test metric
    
    \STATE Compute the maximum number of brackets:
    $ s_{\text{max}} = \left\lfloor \log_\eta (epochs_{\text{max}}) \right\rfloor $
    
    \FOR{ each bracket \( s = s_{\text{max}}, s_{\text{max}}-1, \ldots, 0 \)}
        \STATE Set the number of configurations $ n_{configs} = \left\lceil \frac{s_{\text{max}}+1}{s+1} \cdot \eta^s \right\rceil $ in this bracket
        \STATE Set the initial number of epochs of training per configuration \( n_{initial} = \frac{epochs_{\text{max}}}{\eta^s} \)
        \STATE Randomly select $n_{configs}$ hyperparameter configurations from $S$ in a set $T$
        \FOR{ \( i = 0, 1, \ldots, s \)}
            \STATE Train each configuration from $T$ for $n_{initial}*\eta ^i$ epochs. Test and save the results.
            \STATE Select the top \( \left\lfloor \frac{n_{configs}}{\eta^i} \right\rfloor \)
            configurations of $T$ based on the test metric, discard the rest from $T$
        \ENDFOR
        \STATE Save the best hyperparameter configurations from $T$ in a set $R$
    \ENDFOR
    \RETURN the set $R$ of best hyperparameters
    
    \end{algorithmic}
\end{algorithm}

Practically implemented in \texttt{KerasTuner} as the Hyperband tuner, it is
available for single or multiple GPUs for parallel tuning. Users can specify
$\eta$ and $epochs_{\text{max}}$. This allows control over the search time while
efficiently exploring the hyperparameter search space~\cite{automl_in_action}.

A disadvantage of the Hyperband method is that it will quickly discard the
architectures that performed badly early on in their training. Sometimes
these architectures would outperform the others that did better early on if
allowed to train until completion.

\section{Methodology}
\label{sec:Estimation}

\subsection{Data preprocessing}
\label{sec:Data}
To circumvent the high
dimensionality of our data ($\sim$1000 features with under $4000$ samples),
the number of features was reduced approximately by a factor
of $3$ by removing features in the prepared datasets `time-derived' from
original features, such as by lagging original features by fixed time
steps or taking the mean over the same time steps. Such time-derived
features are standard in the industry~\cite{CNNpred}, but are fundamentally
used because the algorithms applied to the data can only take in one row of
data as input. All the more sophisticated neural network architectures take
in a sequence of rows of data,
making time-derived features redundant. See Section~\ref{sec:Architectures} for
more details.
The datasets were then normalized before applying principal component
analysis (PCA) to further reduce the number of features.

\subsubsection{Principal component analysis}
PCA is a dimensionality reduction
technique widely used to transform high-dimensional data -- especially when
the features are highly correlated -- into a
lower-dimensional space by identifying the directions
(principal components) along which the data varies most~\cite{GoodfellowDeep}.
Let $X$ be our training data matrix of size
$n_{samples}~\times~n_{features}$.
PCA is an orthogonal linear transformation of the real inner product feature
space that maps the data $X$
to a new coordinate system. In this transformed system, the direction of
maximum variance by some scalar projection of the data aligns with the first
coordinate
(known as the first principal component), the direction of the second-highest
variance aligns with the second coordinate, and so on. Moreover, the
transformed columns of $X$ are linearly uncorrelated with each other.

\begin{figure}[ht]
    \centering
    \includegraphics[width=\columnwidth]{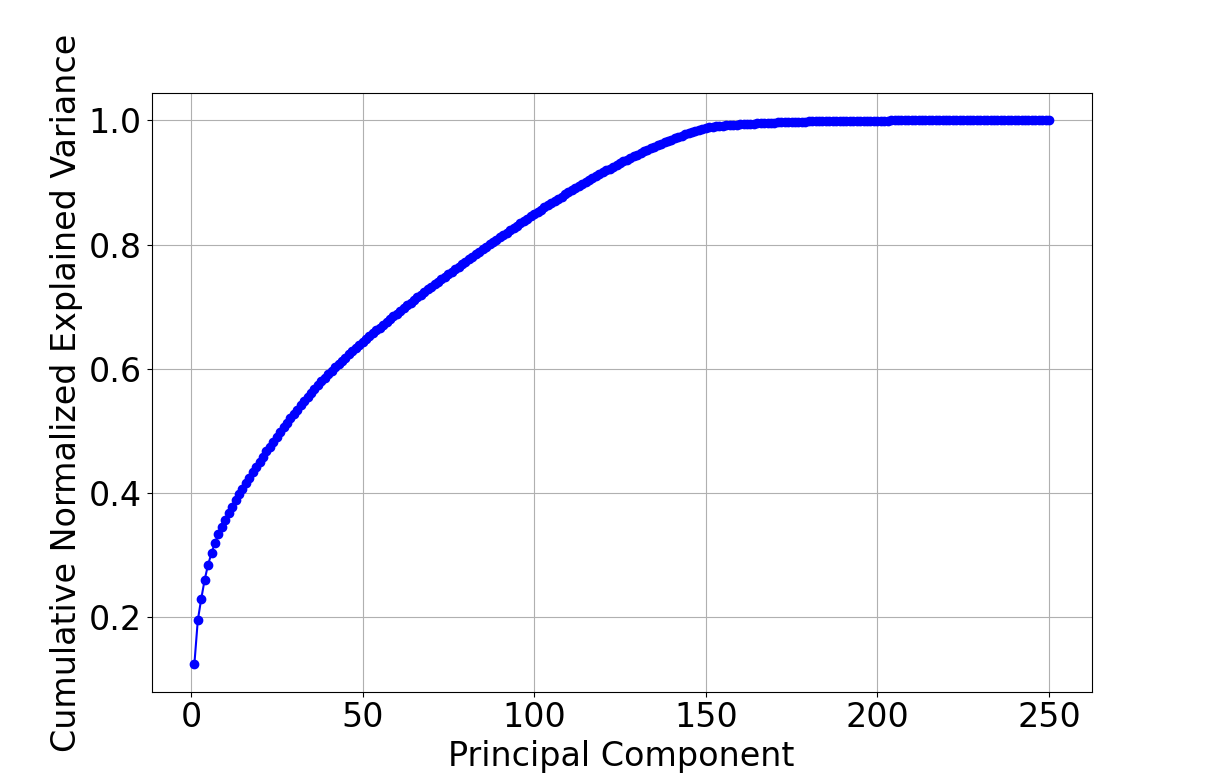}
    \caption{Scree plot for the Japan training dataset after time-derived
    features were removed. The curve saturates around $150$ components,
    suggesting that $k=150$ is a good choice to keep most of the information
    while reducing the dimensionality significantly.}
    \label{fig:scree}
\end{figure}

It turns out that such transformation is given by a matrix $W$ whose
columns are the eigenvectors of
$X^TX$, ordered by their descending eigenvalues, which tell us how much of
the variance in the dataset is described by the corresponding
principal component~\cite{BishopML}. By retaining only the first $k$ components with the highest
variance, i.e., applying a transformation given by the first $k$ columns
of $W$, we transform $X$ into a $n_{samples}~\times~k$ matrix,
effectively capturing most of the information
while reducing dimensionality, noise, and redundancy.
A scree plot is a plot of the cumulative explained variance (i.e., cumulative
sum of the eigenvalues, usually normalized so that the total sum is 1)
against the component number. Looking at the scree
plot, a good value of $k$ that balances dimensionality reduction
and information retention can be identified (see Figure~\ref{fig:scree}
for an example).
We tried a couple values of $k$ for each dataset and architecture type
before proceeding with the main NAS studies.

\subsection{Metrics}
There are many evaluation metrics applicable to the binary classification
problem.
Although common for other classification tasks, accuracy ($ACC$) is not a
good metric as a trivial predictor that always gives the same output no matter
the input can have high accuracy in the case of an unbalanced dataset.
Balanced accuracy ($bACC$), the average between the true negative and true
positive rates, mitigates this problem.
Another popular alternative is the $F_1$ score, the harmonic mean of precision
and recall. The area under the receiver operating characteristic (ROC) curve,
or $AUC$ for short, is another good choice. It is hard to say which one is
better, and a good strategy could mix a combination of these. We mostly
focused on $AUC$ as we found it to be the hardest metric to optimize, but
we paid close attention to $bACC$ and $F1$ too and report on all three
in Section~\ref{sec:Results}. 

\subsection{Random seed variation mitigation}
\label{sec:Random}
Neural networks, in particular in the context of financial time series
forecasting suffer from
comparatively high variance depending on the random seed chosen during
training due to their stochastic
nature~\cite{CNNpred, ClassificationFinanceNN, LSTMForFinance}.
Even identical neural
network architectures, trained in exactly the
same way but with different random seeds can have vastly different performance
on the test dataset, e.g. an $AUC$ of $0.56$ on a `good' random seed and $0.48$
(i.e., worse than random) on a `bad' one. This is a problem for NAS as what
seemed
like a good (or bad) neural network configuration might have just been a lucky
(or unlucky, respectively) seed. To mitigate this problem, we modified
Algorithms~\ref{alg:Bayesian},~\ref{alg:Reinforcement}, and~\ref{alg:Hyperband}
to train and
test each architecture configuration chosen $15$ times with different random
seeds, averaging the test metrics and using the averages as the result for
the configuration chosen. At the
end of optimization, the best performing configuration was trained and tested
$50$ times (again, with different random seeds every time) to further eliminate random
seed variation.

\subsection{Experimental setup}
We started with  $7$ different broad architecture types: FFNN, depthwise 1D CNN,
ordinary 1D CNN, 2D CNN, simple RNN, LSTM, and TFT, but for final optimization studies
only FFNN, ordinary 1D CNN, and LSTM were used, as discussed in
Section~\ref{sec:Architectures}. Each of them has their own search space of
hyperparameters determining the exact network architecture. 

The data was split into (in historical
order): first $70\%$ for training, the following $20\%$ for validation
used in the evaluation during optimization runs,
and the last $10\%$ for final testing on unseen data. Normalization and
PCA transformations were fitted on the training datasets and applied to
training, validation, and test data.

Each of the three search spaces was explored by each of the three search
strategies described in Section~\ref{sec:Strategies} (modified by repeated
training, testing, and averaging as discussed above) on each of the three
(modified) datasets, resulting in $27$ \emph{studies} total.

Each study was limited to 300 \emph{trials}, i.e., 300 different neural
network architecture configurations from a search space tested. Each
configuration was trained for 80
epochs~\footnote{$epochs_{\text{max}}=80$ in the case of Hyperband.},
15 times over (see Section~\ref{sec:Random} above).

When the optimization
studies were completed, the 27 architectures performing best on the validation
datasets were selected and evaluated on the test set. For these final tests
we retrained and tested each network 50 times and averaged the test metrics
to minimize the variance coming from different random seeds.

\section[Results]{Results}
\label{sec:Results}
Our best results were obtained on the Germany dataset, with time-derived
redundant features removed but no PCA applied. The LSTM with parameters
selected by the
hyperband method applied to the unseen test data achieved an $AUC$ score of
$0.56$ on average over $50$ test runs starting from different random seeds
(see Section~\ref{sec:Estimation}), with $0.05$ standard deviation. The same
architecture showed balanced accuracy score of $0.54\pm 0.04$. 1D CNNs
were also able to achieve good performance on this dataset, with the best
architecture provided
by Bayesian optimization giving an $AUC$ score of $0.54\pm 0.05$ and a high $F1$
score of $0.6 \pm 0.06$.

For the Japan dataset, the redundant time-derived features removed
and PCA applied retaining most of the information in the dataset while
further cutting the number of features approximately in half gave the best
results. The best
performing architecture was a 1D CNN coming from Bayesian
optimization, achieving an $AUC$ score of $0.54\pm  0.03$ over $50$ test runs,
with a high $F1$ score of $0.65 \pm 0.02$.

The US dataset was most challenging for us. Although tuned 1d CNNs gave an
AUC score of $0.6\pm 0.02$ on validation data (not used for training) in our
repeated testing, and an individual lucky random seed gave
an $AUC$ of $0.58$ on test data, on average no architecture could achieve $AUC$ 
over $0.5$ on
the test dataset. We believe something very significant happened in the US
market in the last year or so of our data that changed the market dynamics
completely.

\subsection{Architectures and search strategies compared}
Our results show LSTMs and 1D CNNs outperforming simple feedforward networks
for nearly all search strategies and datasets, as expected, but were neck
in neck between each other. 

\begin{figure}[ht]
    \centering
    \includegraphics[width=\columnwidth]{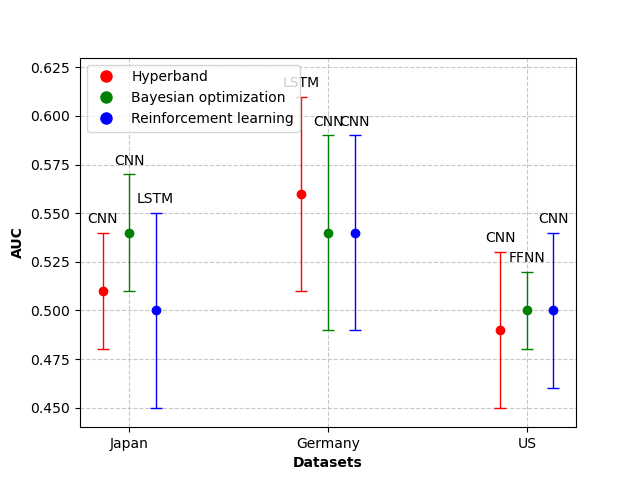}
    \caption{Best performing architectures selected by each search strategy on
    each dataset.
    Every point represents average $AUC$ score and standard deviation on test
    data after retraining the selected architecture $50$ times. Type of the
    neural network chosen by the search strategy is displayed above each point;
    search strategies are color-coded.}
    \label{fig:results}
\end{figure}

For search strategy, overall the hyperband method and Bayesian optimization
showed better performance than the reinforcement learning based approach,
but all three were very close to each other, well within
standard deviation as shown in Figure~\ref{fig:results}.
Optimization time
was very similar between the three methods, about 12 hours per architecture
type using a single Nvidia Quadro RTX 4000 GPU, with feedforward networks
training faster than CNNs and LSTMs due to their lower complexity.

Practically, the
reinforcement learning approach required the most work to adapt from the
open-source implementation of the method in~\cite{Zoph}, and has many
parameters of its own to be optimized for best performance. In contrast,
both the hyperband implementation in \texttt{KerasTuner} and the
TPE Bayesian
optimization from \texttt{Optuna} are easy to use and well-supported,
providing many options for both optimization itself and performance
monitoring. The
\texttt{Optuna} implementation is flexible in that nearly the same
code can be used to optimize hyperparameters of a different kind of solver
other than neural networks, such as decision trees-based algorithms popular
for time series prediction. Meanwhile, \texttt{KerasTuner} implementation of
hyperband has advantage of easy parallelization over several GPUs (although
we did not have the hardware to make use of it).

\section{Discussion}
\label{sec:Discussion}
\begin{figure}[ht]
    \centering
    \includegraphics[width=\columnwidth]{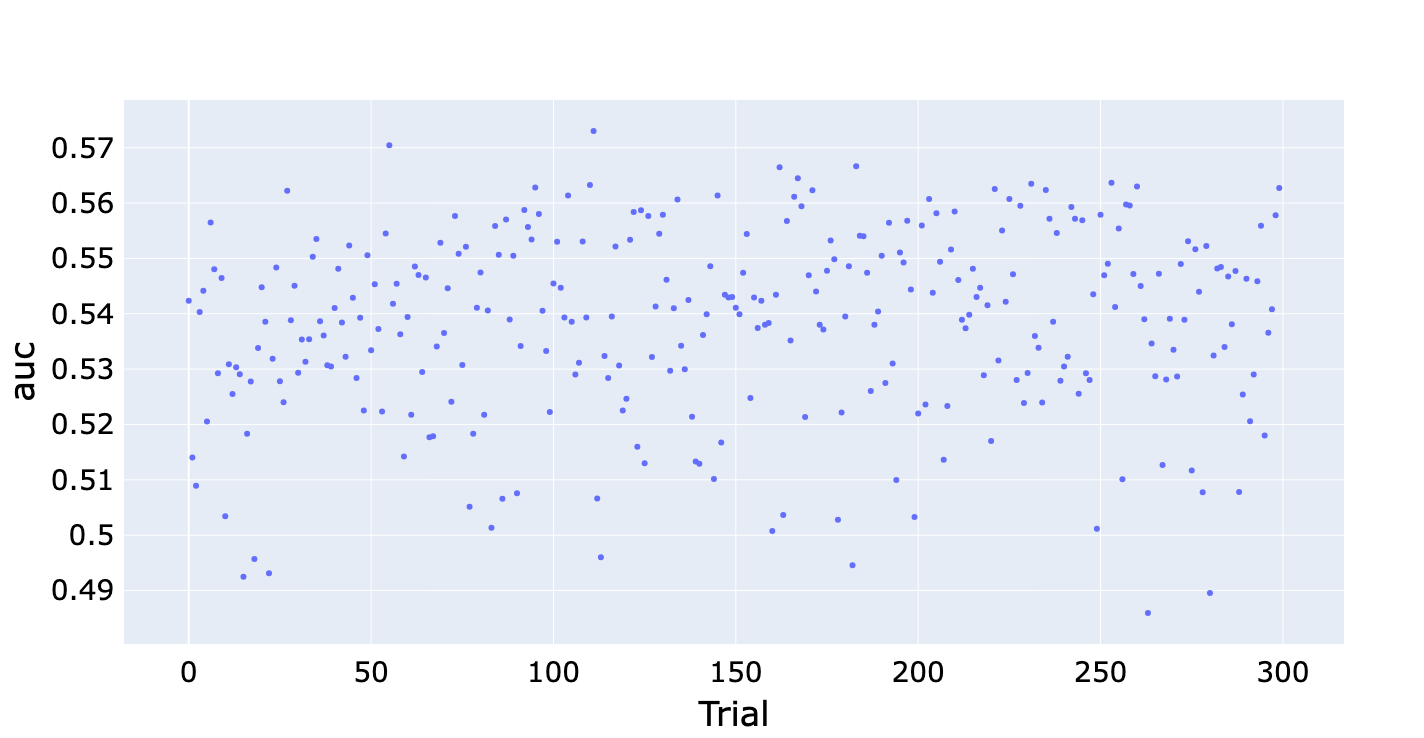}
    \caption{Bayesian optimization history for the ordinary 1D CNN architecture
    on the US dataset. Each point represents average $AUC$ score on the
    validation dataset over $15$ runs for the same network configuration.}
    \label{fig:optim_history}
\end{figure}

Overall, we did not observe significant convergence over time in the
hyperparameter optimization. An example history
of optimization plot 
is shown in Figure~\ref{fig:optim_history}. This was a general problem
present for all neural architecture types, optimization strategies and datasets.

We believe all search strategies were working as they should, but the main
problem of financial time series prediction on our datasets was just too
difficult to decisively improve upon through neural architecture search alone.
Many steps were taken to improve overall performance. In the beginning the
networks were heavily overfitting, where we struggled to get any metrics
above $0.5$ even on the validation data. The feature reduction measures such
as removing time-derived features and PCA discussed in Section~\ref{sec:Data}
helped mitigate this. Unfortunately, this still did not always translate to
good performance on the test dataset, even though we heavily used dropout
layers for regularization. We also tried converting the main binary
classification problem to regression, but this did not result in substantial
improvement.

As a consequence, simply taking the final suggested architecture by either
search strategy did not give the best results. The best performing ones were
also not always the best choice due to the random seed variance discussed
in Section~\ref{sec:Random}. Even though each network configuration was trained
and tested $15$ times during the optimization process, training and testing the
best architectures more (e.g., 50) times sometimes showed worse
performance, indicating that the issue of random seed variation was not fully
resolved in our approach.

\begin{figure}[ht]
    \centering
    \includegraphics[width=\columnwidth]{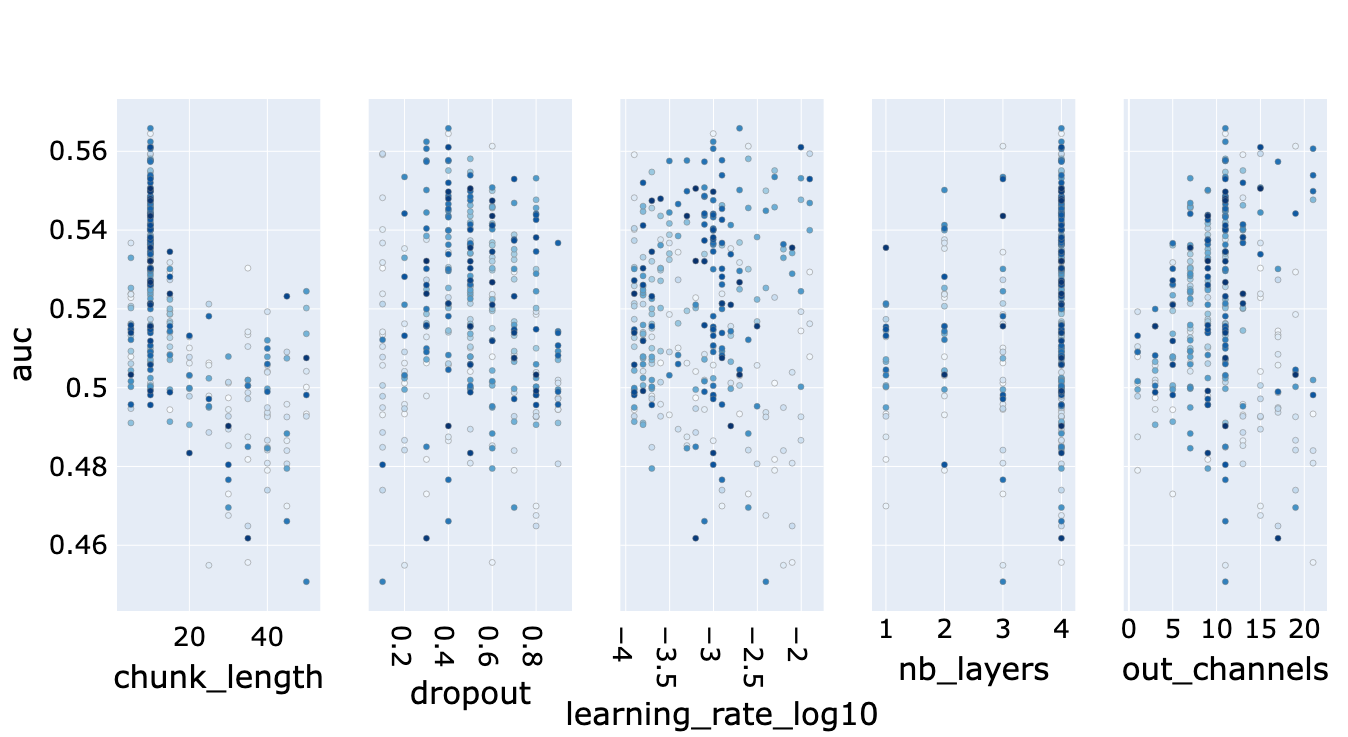}
    \caption{Slice plot for Bayesian optimization of LSTMs on the Japan dataset.
    Each point represents average $AUC$ score on the validation dataset over
    $15$ runs for the same network configuration. It is clear that setting
    \texttt{chunk\_length} (i.e. the length in time of a sequence passed to the
    LSTM for individual prediction) to $10$ gives the best results, while the
    other parameters are less impactful.}
    \label{fig:japan_lstm_slice_auc}
\end{figure}

Nevertheless, all search strategies were still successful in highlighting
best choices for network parameters, at least for the validation set,
which is the most evident from slice plots as
in Figure~\ref{fig:japan_lstm_slice_auc}. For Bayesian optimization and the
reinforcement learning approach, the best
parameters for each architecture were read off slice plots like this.

To combat the big variance between networks starting from different random seeds
(see Section~\ref{sec:Data} and Figure~\ref{fig:results}),
we tried filtering the outputs,
disregarding those with probabilities close to $0.5$ (interpreted as those
where the network `was not sure'), a standard approach~\cite{GoodfellowDeep}.
Although this improved overall performance a little on average, it instead
greatly increased the variance. For a fixed
architecture, selecting networks trained from random seeds giving good
performance on validation set did not translate to better performance on the
test set. Our best suggestion to combat the variance is to use an ensemble model
built from many incarnations of a chosen network architecture.

\subsection{Future work}

\label{sec:Future}
Performing the same studies on more and publicly available time-series
datasets, especially non-financial data would shed more light on whether our
findings were specific to our data or a more general feature.

In addition, cell-based and hierarchical search
spaces~\cite{White_NAS_survey} could be explored in the future in the context
of time-series forecasting. These are based on searching over structural blocks
within a network piece by piece instead of navigating through a grid-like space
as in the standard approach we took.
One-shot NAS techniques such as
supernet-based methods (differentiable and not) and
hypernetworks~\cite{White_NAS_survey} could be explored. With one-shot
techniques, one trains a single (massive) super/hyper-network, which can be
used to subsample and evaluate many smaller networks to find the optimal one,
without re-training.
This is opposed to the standard approach of training and testing many
architecture configurations that we took. Finally, genetic algorithms have
proved to be very successful for NAS on image datasets~\cite{GeneticNAS} and
should be explored in the context of time series too.

\section*{Statements and Declarations}

\begin{itemize}
\item Funding

Funded by InnoSuisse (Project number: 54228.1 IP-ICT).
First author was also supported by HEIG-VD projects HES-SO-340375,
HES-SO-117951, and HES-SO-122845.

\item Competing interests

The authors declare no competing interests.
\item Data availability statement

The datasets presented in this article are private customer data provided
by project partner Predictive Layer SA and are thus not readily available.

\item Author contributions

Conceptualization, Stephan Robert-Nicoud, Biagio Nigro and Denis Levchenko;
Methodology, Denis Levchenko;
Software, Denis Levchenko, Biagio Nigro, Shabnam Ataee and Efstratios Rappos;
Validation, Denis Levchenko, Efstratios Rappos and Shabnam Ataee;
Investigation, Denis Levchenko, Biagio Nigro, Efstratios Rappos and Shabnam Ataee;
Resources, Biagio Nigro and Stephan Robert-Nicoud;
Writing – Original Draft Preparation, Denis Levchenko, Biagio Nigro, Shabnam Ataee, Efstratios Rappos;
Writing – Review and Editing, Denis Levchenko, Efstratios Rappos; Supervision, Stephan Robert-Nicoud;
Project Administration, Biagio Nigro and Stephan Robert-Nicoud;
Funding Acquisition, Stephan Robert-Nicoud.
\end{itemize}

\bibliography{refs}
\bibliographystyle{ieeetr}

\end{document}